\documentclass[aps,longbibliography,prd,groupedaddress]{revtex4}
\usepackage{graphicx}
\usepackage{amsmath,amssymb,amsfonts}
\usepackage[colorlinks,citecolor=blue,linkcolor=blue,urlcolor=blue]{hyperref}

\begin{document}

\begin{center}

{\ }
\vspace{2cm}

{\large\bf White-hole dark matter and the origin of past low-entropy}\\[5mm]

\vspace{2cm}
{Francesca Vidotto}\\[3mm]
{University of the Basque Country UPV/EHU, Departamento de F\'isica Te\'orica, 
Barrio Sarriena, 48940 Leioa, Spain\\
francesca.vidotto@ehu.es}

\vspace{1cm}
{Carlo Rovelli}\\[3mm]
{CPT, Aix-Marseille Universit\'e, Universit\'e de Toulon, CNRS, case 907, Campus de Luminy, 13288 Marseille, France\\
{rovelli@cpt.univ-mrs.fr}}\\[1cm]

{Submission date April 1, 2018}

\end{center}

\vspace{2cm}

{\bf Abstract}

\vspace{.5cm}

\noindent Recent results on the end of black hole evaporation give new weight to the hypothesis that a component of dark matter could be formed by remnants  of evaporated black holes: stable Planck-size white holes with a large interior. The expected lifetime of these objects is consistent with their production at reheating. But remnants could also be pre-big bang relics in a bounce cosmology, and this possibility has strong implications on the issue of the source of past low entropy: it could realise a perspectival interpretation of past low entropy. The ideas briefly presented in this essay are developed in forthcoming papers.

\vspace{6cm}
\begin{center}
\underline{\it Essay written for the Gravity Research Foundation 2018 Awards for Essays on Gravitation.}
\end{center}

\newpage

\section{Remnants}

The possibility that remnants of evaporated black holes form a component of dark matter was suggested by MacGibbon \cite{J.H.MacGibbon1987} and has been explored in \cite{Barrow1992, Carr:1994ar, Liddle1997, Alexeyev2002,Chen2003, Barrau:2003xp, Chen2004, Nozari2008}. There are no strong observational constraints on this potential contribution to dark matter \cite{Carr2016} and the weak point of the scenario has been, so far, the question of the physical nature of the remnants. 

Here we point out that: (i) the picture has  changed  because of the realisation that conventional physics provides a candidate for remnants: small-mass white holes with large interiors \cite{Rovelli2014h,Haggard2014,DeLorenzo2016}, which can be stable if they are sufficiently light \cite{Bianchi2018}, and that  \emph{are} produced at the end of the evaporation as recent results in quantum gravity indicate \cite{Christodoulou2016,Christodoulou2018, DAmbrosio2018,Rovelli2018a};  (ii) the remnant lifetime predicted in \cite{Bianchi2018} happens to be consistent with the production of primordial black holes at the end of inflation; and (iii) even more interesting, if remnants come from a pre-big-bang phase in a bouncing cosmology, they provide a solution to the problem of the early-universe low-entropy, and hence to the problem of the arrow of time \cite{Rovelli:2015}.

\section{White holes}

The difference between a black hole and a white hole is not very pronounced.  Observed from the outside (say from the exterior of a sphere of radius $r=2m+\epsilon>2m$, where $m$ is the mass of the hole)  for a finite amount of time, a white hole cannot be distinguished from a black hole:   
Zone II of the maximally extended Schwarzschild solution is equally the outside of a black hole and the outside of a white hole (see FIG. 1, Left). Analogous considerations hold for the Kerr solution. 

\begin{figure}[h]
	\includegraphics[width = .33 \columnwidth]{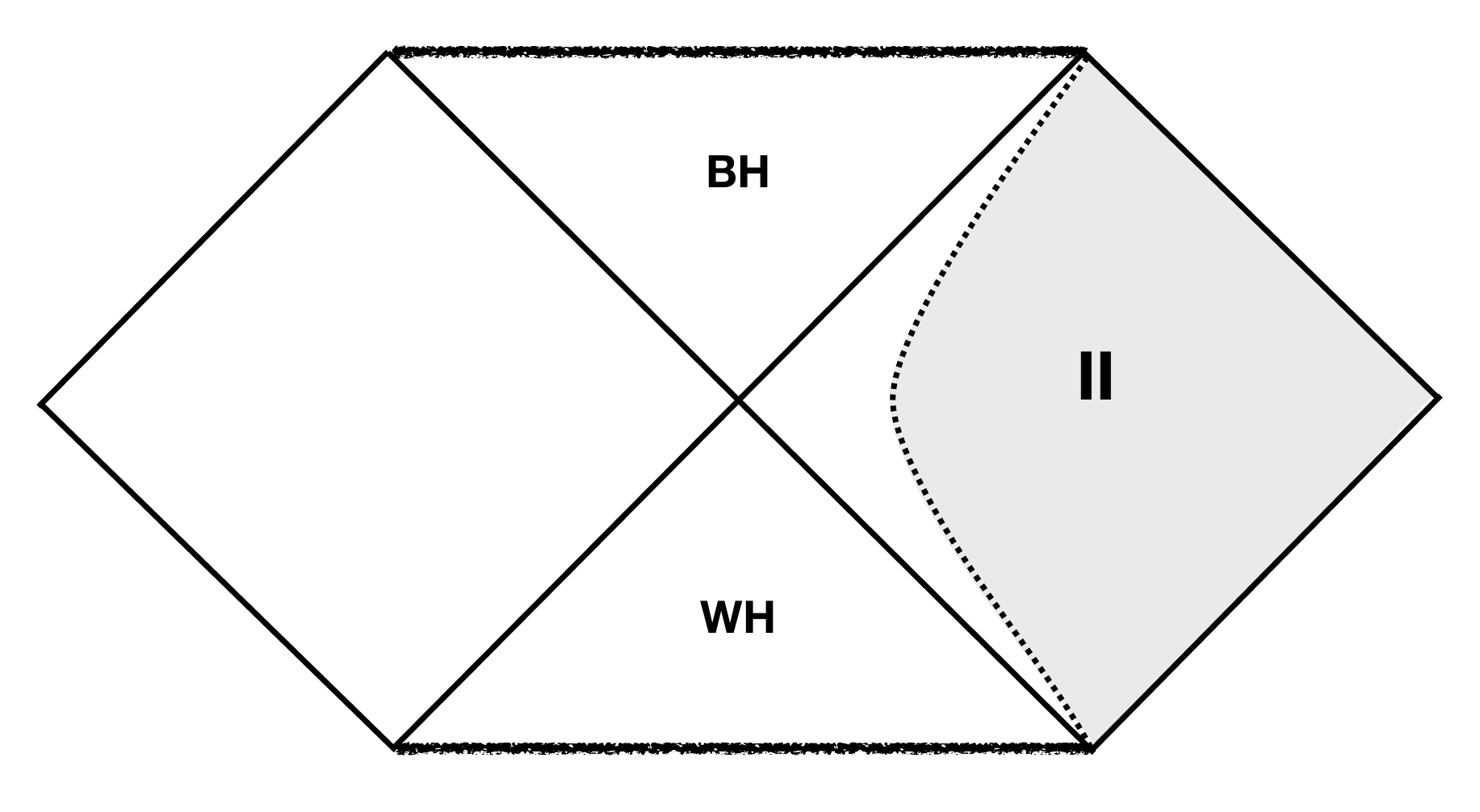}\hfil
	\includegraphics[width = .33 \columnwidth]{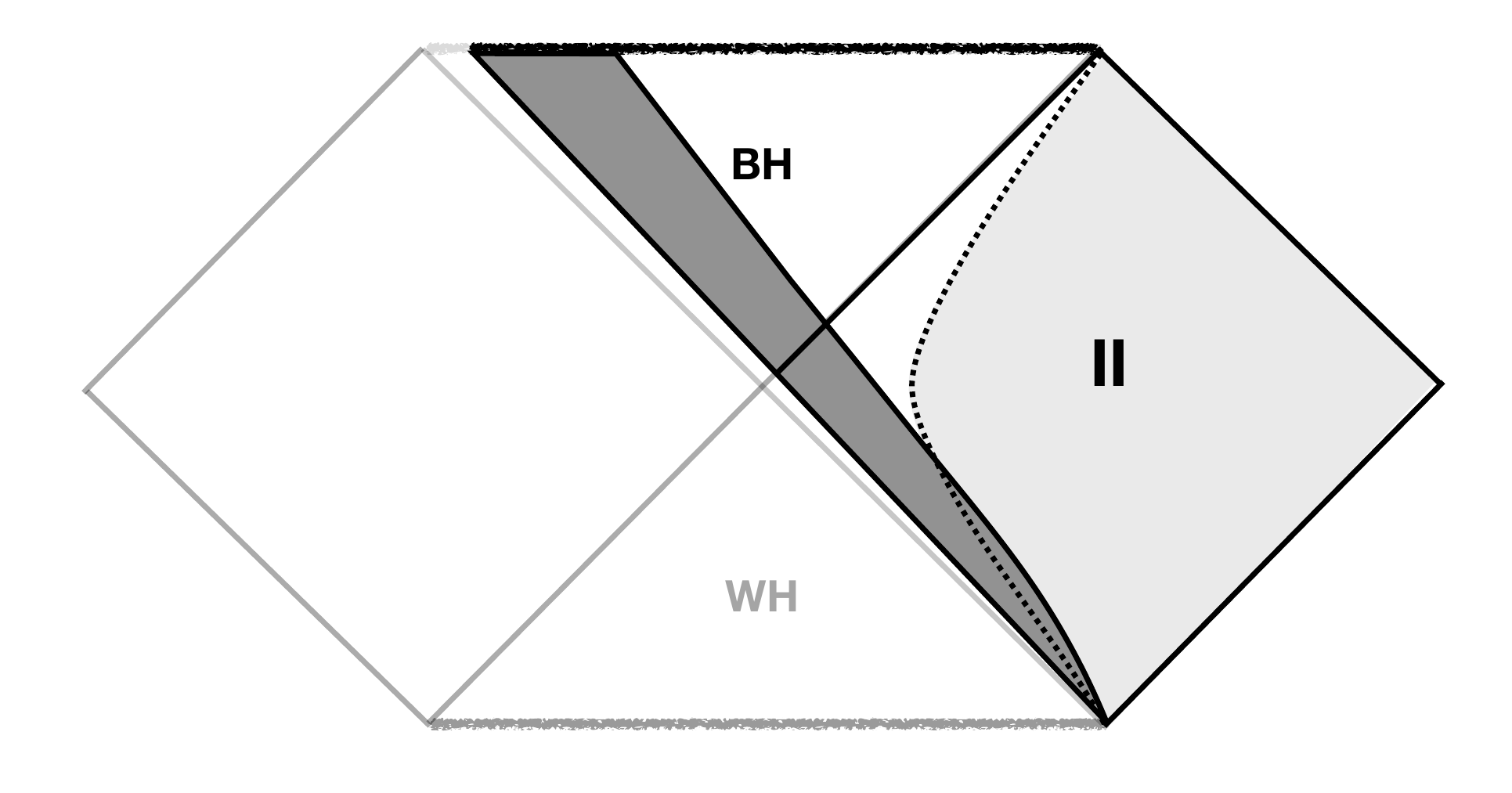}\hfil
	\includegraphics[width = .33 \columnwidth]{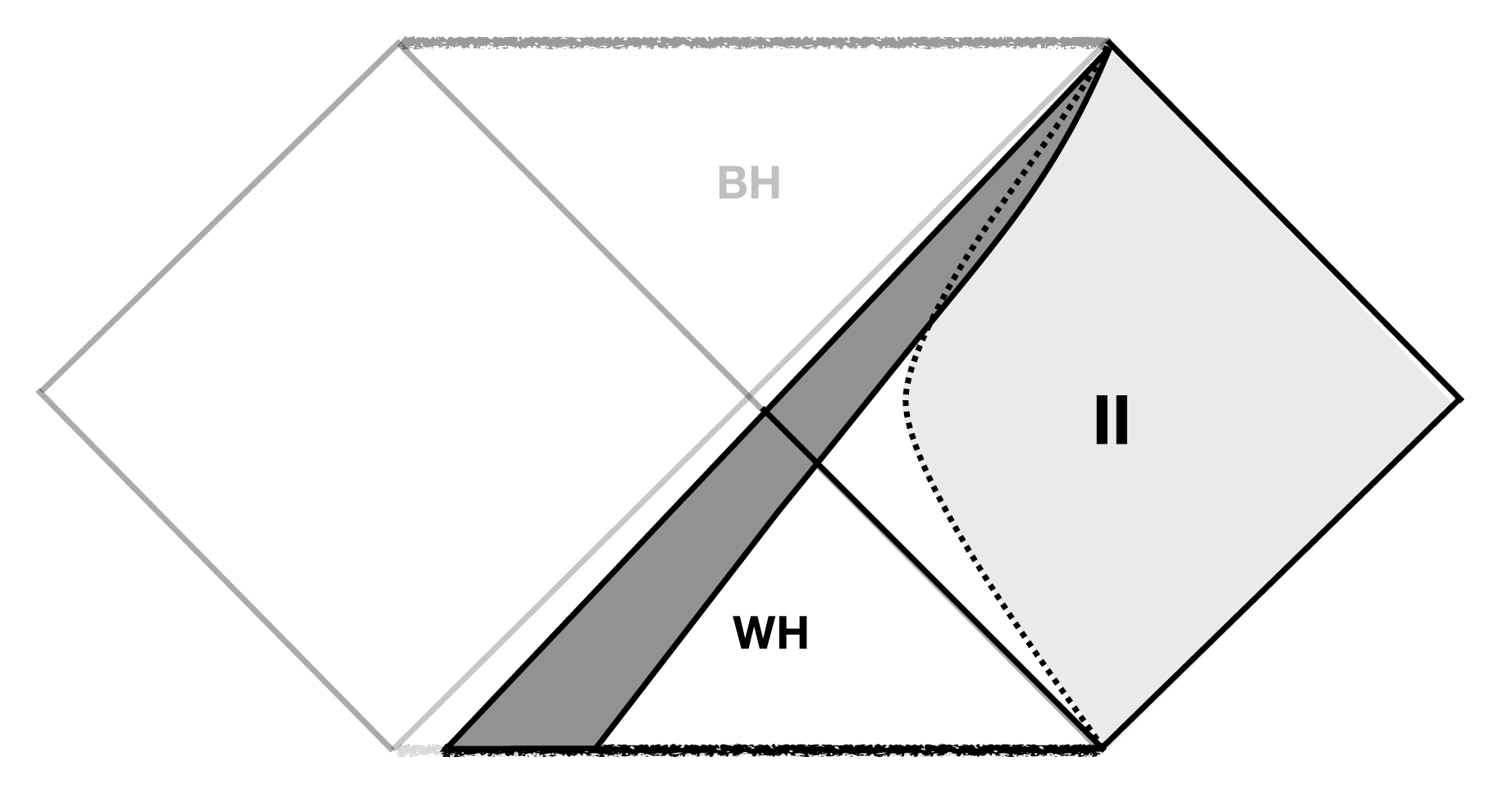}
\caption{\em Left: Extended Schwarzschild spacetime. The (light grey) region outside $r=2m+\epsilon$ is equally the outside of a black and a white hole. Center: A collapsing star (dark grey) replaces the white hole region in the non-stationary collapse metric.  Right: The time revered process. The difference between the last two is only in the far past or future.}
	\label{Kruskal}
	\end{figure}
	
The objects in the sky we call `black holes' are approximated by a stationary metric only for a limited time.   At least in the past the metric was non-stationary, as they were produced by gravitational collapse. The continuation of the metric inside the radius $r=2m+\epsilon>2m$ contains a trapped region, but not an anti-trapped region. Instead of an anti-trapped region there is a collapsing star.  Thus from the outside a `black hole' (as opposed to a `white') is only characterised by not having the anti-trapped region in the past (see FIG. 1, Center). Vice versa, from the exterior and for a finite time, a white hole is indistinguishable from a black hole, but in the future ceases to be stationary and there is no trapped region (see FIG.1, Right). 

The classical prediction that the black is stable forever is not reliable.  In the uppermost band of the central diagram of FIG. 1 quantum theory dominates.  In other words, the death of a black hole is a quantum phenomenon. The same is true for a white hole, reversing the time direction. That is, the birth of a white hole is in a region where quantum gravitational phenomena are strong.  This consideration eliminates a traditional objection to the physical existence of white holes: How would they originate? The answer is that they originate from a region where quantum phenomena dominate the behaviour of the gravitational field.  

Such regions are generated in particular by the end of the life of a black hole.  Hence a white hole can be originated by a dying black hole.  This has been shown to be possible in \cite{Haggard2014} and was explored in \cite{Christodoulou2016,DeLorenzo2016,Christodoulou2018,Bianchi2018, DAmbrosio2018, Rovelli2018a}. Thus the life cycle of the black-white hole is given in FIG. 2. 

\begin{figure}[h]
	\includegraphics[width = .15 \columnwidth]{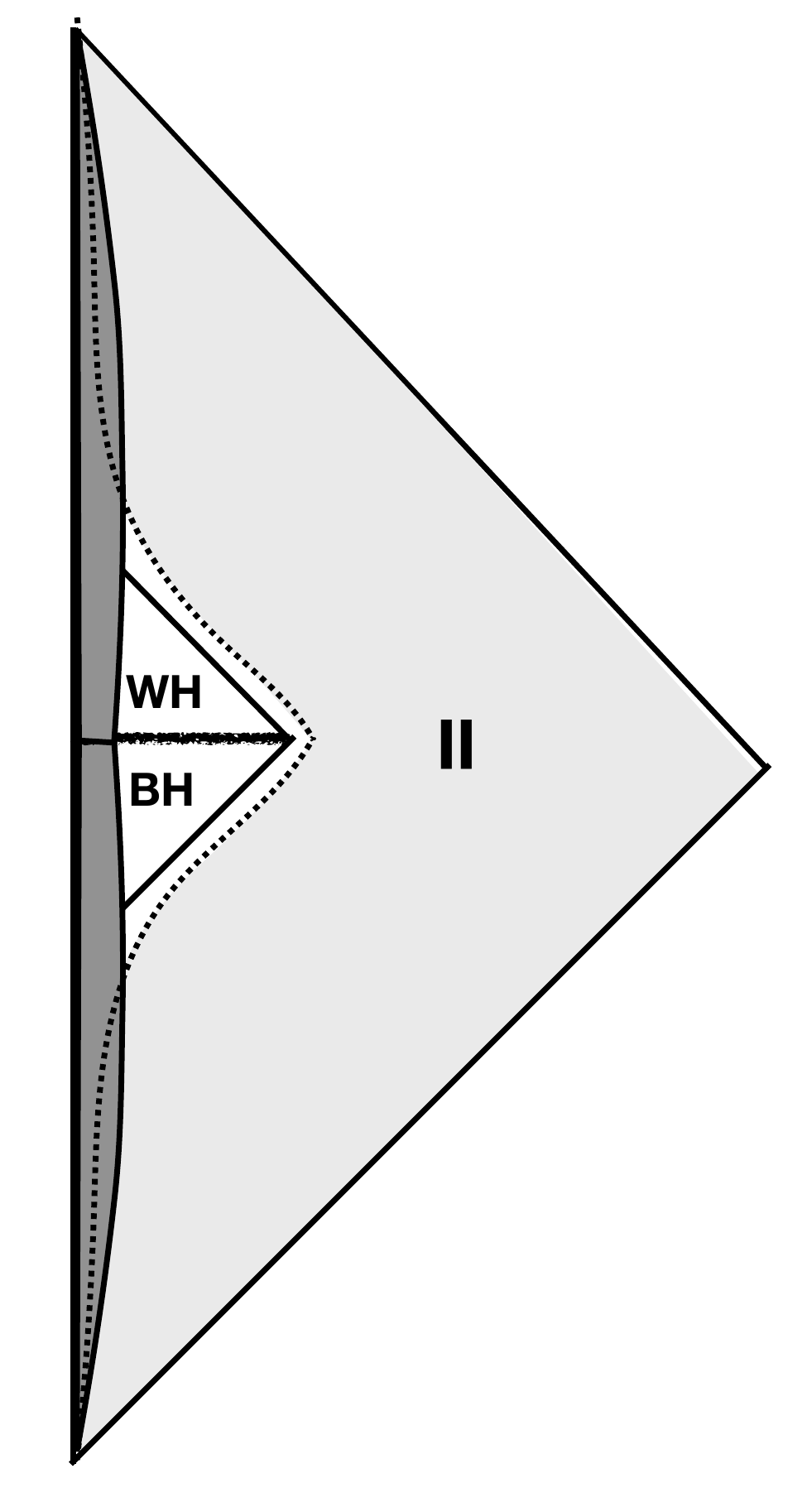}
\caption{\em The full life of a black-white hole.}
	\label{bwh}
	\end{figure}
	
The process is asymmetric in time \cite{DeLorenzo2016,Bianchi2018} and the time scales are determined by the initial mass of the hole $m_o$.  The lifetime $\tau_{BH}$ of the black hole is known via Hawking radiation theory 
\begin{equation}
\tau_{BH}\sim m_o^3
\end{equation}
in Planck units $\hbar=G=c=1$.  
This time can be as short as $\tau_{BH}\sim m_o^2$ because of quantum gravitational effects \cite{Rovelli2014h,Haggard2014} (see also \cite{Gregory:1993vy,Casadio:2000py,Casadio:2001dc,Emparan:2002jp,Kol:2004pn}) but we disregard this possibility here.
\\
The lifetime of the white hole phase, $\tau_{W\!H}$, is longer \cite{Bianchi2018}:
\begin{equation}
\tau_{W\!H}\sim m_o^4. 
\end{equation}
The tunnelling process itself takes a time of the order of the current mass \cite{Christodoulou2018,Barcelo2016}.
\begin{equation}
\tau_{T} \sim m. 
\end{equation}

The classical evolution and the \emph{exterior} of a (stationary, spherically symmetric) black hole is uniquely determined by its current horizon area, but its \emph{internal} properties are not. The internal volume keeps increasing with time \cite{Christodoulou2015,Bengtsson2015,Ong2015,Wang2017, Christodoulou2016a} and determines post tunnelling evolution.  Therefore the state of the hole at a given  time is not specified solely by its current mass $m$, but also by its internal geometry, in turn determined by $m_o$. We write the quantum state of the hole in the form $|m_o,m\rangle_B$, where the first quantum number is the initial mass which determines the interior, the second is the current mass, which determines the horizon area. Here is  the life cycle of a collapsed object
\begin{eqnarray}
\xrightarrow[\text{collapse}]{} |m_o,m_o\rangle_B  \xrightarrow[\text{black hole}]{\tau_{W\!H}\sim m_o^3}
 |m_o,m_{P\ell}\rangle_B  \xrightarrow[\text{tunnelling}]{\tau_{T}\sim m_{P\ell}}
  |m_o,m_{P\ell}\rangle_W \xrightarrow[\text{white hole}]{\tau_{W\!H}\sim m_o^4} 
|m_{P\ell},m_{P\ell}\rangle_W  \xrightarrow[\text{end}]{ }. 
\end{eqnarray}

\section{Stability}

Large classical white holes are unstable under perturbations \cite{Frolov2012}. The wavelength of the perturbation needed to trigger the instability must be smaller that the size of the hole. To make a Planck size white hole unstable, we need trans-Planckian radiation, and this is not  allowed by quantum gravity  \cite{Bianchi2018}.  Even independently from this, if a Planck-scale white hole is unstable, its decay mode is into a Planck-scale black hole:
\begin{equation}
 |m_o,m_{P\ell}\rangle_W  \xrightarrow[\text{instability}]{}  |m_o,m_{P\ell}\rangle_B , \label{inst}
 \end{equation}
which is equally unstable to tunnel back into a white hole 
\begin{equation}
 |m_o,m_{P\ell}\rangle_B  \xrightarrow[\text{tunnelling}]{ }  |m_o,m_{P\ell}\rangle_W . \label{tunn}
\end{equation}
Since the two are indistinguishable from the exterior, this oscillation has no effect on the exterior: the Planck size remnant remains a Planck size remnant. The possibility of this oscillation has been considered in \cite{Barcelo:2015uff,Barcelo2016,Garay2017}.   

\section{Cosmology: reheating}

White-hole remnants can be a constituent of dark matter.  A local dark matter density of the order of $0.01 \, M_\odot\!/pc^3$ corresponds to approximately one Planck-scale white hole per each $10.000 \, K\!m^3$.  

For these objects to be present now we need  their lifetime 
to be larger than or equal to
 the Hubble time $T_H$ 
\begin{equation}
                               m_o^4\ge T_H. 
\end{equation}
On the other hand, the lifetime of the black hole phase must be shorter than the Hubble time  
\begin{equation}
                               m_o^3  <  T_H. 
\end{equation}
This gives the possible value of $m_0$:
\begin{equation}
                             10^{10} gr  \le m_o < 10^{15} gr.
\end{equation}
Their Schwarzschild radius is in the range  
\begin{equation}
                             10^{-18} cm  \le R_o < 10^{-13} cm.
\end{equation}
Black holes of a given mass could have formed when their Schwarzschild radius was of the order of the horizon. Remarkably, the horizon was presumably in this range at the end of inflation, during or just after reheating. This concordance supports the plausibility of the scenario. 

\section{Erebons and the arrow of time}

Roger Penrose has coined the name \emph{erebons}, from the Greek god of darkness Erebos, to refer to matter crossing over from one eon to the successive one \cite{Penrose2017} in cyclic cosmologies \cite{Penrose2012}.  Remnants could also be formed in a contracting phase before the current expanding one  \cite{Quintin2016,Carr2017} in a big bounce scenario \cite{Brandenberger2016,Agullo2016}.  

A surprising aspect of the white-hole erebons scenario is that it addresses the low-entropy of the initial cosmological state. 

The current arrow of time can be  traced back to the homogeneity of the gravitational field at the beginning of the expansion: this homogeneity is the  reserve of low-entropy that drives current irreversibile phenomena \cite{Penrose1979a}.  If the big bounce was actually finely dotted by holes with large interior, the metric was \emph{not} homogeneous.  Its low entropy is a consequence of the fact that we do not access the vast interiors of these holes.  In other words, it is determined by the peculiar subset of observables to which we happen to have access.

This represents a realisation of the perspectival interpretation of entropy \cite{Rovelli:2015}.  Entropy depends on a microstate of a system {\em and a choice of macroscopic observables}, or coarse graining.   For any microstate there is some choice of macroscopic observable that make entropy low.  Early universe could have low entropy  not because its microstate is peculiar, but because the coarse graining under which we access it is peculiar.  Past low entropy, and its consequent irreversible evolution can be real, but perspectival, like the apparent rotation of the sky.

If the cosmos at the big bounce was finely dotted with white holes with large interiors, the gravitational field was not in an improbable homogeneous configuration.  It was in a generic crumpled configuration. But being \emph{outside} all white holes, we find ourselves in a special place, and from the special perspective of this place we see the universe under a coarse graining which defines an entropy that was low in the past.

\vskip1cm
\noindent {\bf Acknowledgements} 
The work of FV at UPV/EHU is supported by \\ the grant IT956-16 of the Basque Government and by the grant FIS2017-85076-P (MINECO/AEI/FEDER, UE).
\vfill

\providecommand{\href}[2]{#2}\begingroup\raggedright\endgroup

\end{document}